\documentclass[twocolumn,final,aps,prb,showpacs,superscriptaddress,amsmath,amssymb,amsfonts,floatfix]{revtex4-1}
\usepackage[utf8]{inputenc}
\usepackage[T1]{fontenc}        
\usepackage{lmodern}            
\usepackage{graphicx}
\usepackage{multirow}
\usepackage{epsfig}
\usepackage{url}
\usepackage{color}

\usepackage[colorlinks,breaklinks,bookmarks=true,citecolor=blue,linkcolor=blue,urlcolor=blue]{hyperref}
\usepackage{color}
\usepackage{tabularx}
\usepackage{sidecap}
\usepackage{float}

\usepackage{orcidlink}

\usepackage{amsmath}            
\usepackage{amstext}            
\usepackage{amssymb}            
\usepackage{mathtools}          
\usepackage{scalerel,stackengine} 
\usepackage{ulem}

\newcommand{\freq}{^{\omega \nu \nu'}}
\newcommand{\operator}[1]{\hat{#1}}

\newcommand{\magn}{\text{m}}

\newcommand{\iu}{{i\mkern1mu}}

\newcommand{\dga}{\textrm{D}\Gamma\textrm{A}}

\newcommand{\torder}{\mathcal{T}}
\newcommand{\annihilation}[1]{\operator{c}_{#1}^{\phantom{\dagger}}}
\newcommand{\creation}[1]{\operator{c}_{#1}^\dagger}
\newcommand{\expec}[1]{\langle {#1} \rangle}
\newcommand{\ph}{\textrm{ph}}
\newcommand{\phb}{\overline{\textrm{ph}}}
\newcommand{\pp}{\textrm{pp}}
\newcommand{\bk}{\textbf{k}}

\begin{document}

\title{Spin fluctuations sufficient to mediate superconductivity in nickelates}

 \author{Paul Worm\,\orcidlink{0000-0003-2575-5058}}
 \affiliation{Institute of Solid State Physics, TU Wien, 1040 Vienna, Austria}

\author{Qisi Wang\,\orcidlink{0000-0002-8741-7559}}
\affiliation{Physik-Institut, Universit\"at Z\"urich, Winterthurerstrasse 190, CH-8057 Z\"urich, Switzerland}
\affiliation{Department of Physics, The Chinese University of Hong Kong, Shatin, Hong Kong, China}

\author{Motoharu Kitatani\,\orcidlink{0000-0003-0746-6455}}
\affiliation{Department of Material Science, University of Hyogo, Ako, Hyogo 678-1297, Japan}
\affiliation{RIKEN Center for Emergent Matter Sciences (CEMS), Wako, Saitama, 351-0198, Japan}

\author{Izabela Bia\l{}o\,\orcidlink{0000-0003-3431-6102}}
\affiliation{Physik-Institut, Universit\"at Z\"urich, Winterthurerstrasse 190, CH-8057 Z\"urich, Switzerland}
\affiliation{AGH University of Science and Technology, Faculty of Physics and Applied Computer Science, 30-059 Krak\'ow, Poland}

\author{Qiang Gao}
\affiliation{Beijing National Laboratory for Condensed Matter Physics, Institute of Physics, Chinese Academy of Sciences, Beijing 100190, China}

\author{Xiaolin Ren}
\affiliation{Beijing National Laboratory for Condensed Matter Physics, Institute of Physics, Chinese Academy of Sciences, Beijing 100190, China}

\author{Jaewon Choi}
\affiliation{Diamond Light Source, Harwell Campus, Didcot OX11 0DE, United Kingdom}

\author{Diana Csontosov\'a}
\affiliation{Department of Condensed Matter Physics, Faculty of
  Science, Masaryk University, Kotl\'a\v{r}sk\'a 2, 611 37 Brno,
  Czechia}

\author{Ke-Jin Zhou}
\affiliation{Diamond Light Source, Harwell Campus, Didcot OX11 0DE, United Kingdom}

\author{Xingjiang Zhou}
\affiliation{Beijing National Laboratory for Condensed Matter Physics, Institute of Physics, Chinese Academy of Sciences, Beijing 100190, China}

\author{Zhihai Zhu}
\affiliation{Beijing National Laboratory for Condensed Matter Physics, Institute of Physics, Chinese Academy of Sciences, Beijing 100190, China}

\author{Liang Si\orcidlink{0000-0003-4709-6882}}
\email{Corresponding author: siliang@nwu.edu.cn}
\affiliation{School of Physics, Northwest University, Xi'an 710127, China}
\affiliation{Institute of Solid State Physics, TU Wien, 1040 Vienna, Austria}
  
\author{Johan Chang}
\affiliation{Physik-Institut, Universit\"at Z\"urich, Winterthurerstrasse 190, CH-8057 Z\"urich, Switzerland}

\author{Jan M. Tomczak\,\orcidlink{0000-0003-1581-8799}}
\affiliation{Department of Physics, King's College London, Strand, London WC2R 2LS, United Kingdom}
\affiliation{Institute of Solid State Physics, TU Wien, 1040 Vienna, Austria}

 \author{Karsten Held\,\orcidlink{0000-0001-5984-8549}}
 \email{Corresponding author: held@ifp.tuwien.ac.at}
 \affiliation{Institute of Solid State Physics, TU Wien, 1040 Vienna, Austria}

\date{\today}

\begin{abstract}
  Infinite-layer nickelates show high-temperature superconductivity,
  and the experimental phase diagram agrees well with the one simulated within 
  the dynamical vertex approximation (D$\Gamma$A).
  Here, we compare the spin-fluctuation spectrum behind these calculations 
  to resonant inelastic X-ray scattering experiments. The overall agreement
  is good. 
  This independent cross-validation of the strength of spin fluctuations strongly supports the scenario, advanced by D$\Gamma$A, that spin-fluctuations are the mediator of
  the superconductivity observed in nickelates.
\end{abstract}

\maketitle

\section{Introduction}
Contrasting
cuprates~\cite{Bednorz1986},
to the new nickelate superconductors~\cite{li2019superconductivity,Li2020,zeng2020,Osada2020,Zeng2021,pan2021,Osada2021,Wang2022} offers the unique opportunity to understand high-temperature ($T_c$) superconductivity more thoroughly: the two systems are similar enough to expect a common origin of superconductivity, but at the same time distinct enough to pose severe restrictions on any theoretical description.
Structurally, both, nickelate and cuprate superconductors, consist of  Ni(Cu)O$_2$ planes that host the superconductivity. These layers are separated by  buffer layers of, e.g., Nd(Ca) atoms in the infinite-layer compound NdNiO$_2$(CaCuO$_2$). 
Additionally, both Ni and Cu exhibit a nominal 3$d^9$ electronic configuration in the respective parent compound, with a 3$d_{x^2-y^2}$-derived band that is close to half-filling.

Turning to the differences, a 
 major one is
that for cuprates the oxygen 2$p$ bands, that 
strongly hybridize with the Cu 3$d_{x^2-y^2}$ band, are below but close to the Fermi energy. This makes  the parent compound a charge-transfer insulator~\cite{Zaanen1985}, and
 the  Emery model~\cite{Emery1987} the elemental model for cuprates. For nickelates, on the other hand,  these 2$p$ bands are shifted down
relative to the 3$d_{x^2-y^2}$ band which is fixed to the Fermi energy. As a consequence, the oxygen band is now sufficiently far away from the Fermi energy. While there is still the hybridization with the Ni 3$d_{x^2-y^2}$ band,  the oxygen 2$p$ bands do not host holes if nickelates are doped. Instead, however, the rare earth 5$d$ bands are also shifted down (compared to the Ca bands that are above the Fermi energy in the cuprate CaCuO$_2$), now even cross the Fermi energy and  form two electron
pockets around the $\Gamma$ and $A$ momentum points. This is evidenced by density functional theory (DFT) calculations~\cite{Botana2019,Hirofumi2019,jiang2019electronic,Motoaki2019,hu2019twoband,Wu2019,Nomura2019,Zhang2019,Jiang2019,Werner2019,Si2019,Nomura2022,Kitatani2023b} and, experimentally,
by the negative Hall conductivity~\cite{li2019superconductivity,zeng2020} for the infinite-layer compound. In all, this situation
creates a seemingly more complicated multi-band picture already for the undoped parent compound.

However, one of the pockets, the $\Gamma$ pocket, shifts up and even above the Fermi energy either when (i) doping into the superconducting regime or (ii) when replacing Nd by La in DFT+dynamical mean-field theory (DMFT) calculations~\cite{Kitatani2020,Held22} (and Ca$_x$La$_{1-x}$NiO$_2$ shows a very similar phase diagram as  Sr$_x$Nd$_{1-x}$NiO$_2$). Thus it appears unlikely that the $\Gamma$ pocket is the key for superconductivity in nickelates.
The  $A$ pocket, on the other hand, is more stable but it does not hybridize with the Ni 3$d_{x^2-y^2}$ band\footnote{This pocket also vanishes when going from infinite to finite-layer nickelates\cite{Worm2021c}.}. Hence, in Ref.~\onlinecite{Kitatani2020}
 the pockets  were justifiably treated as a passive electron reservoir, largely decoupled from the  Ni 3$d_{x^2-y^2}$ band~\footnote{At larger doping, outside the superconducting dome, the
  Ni 3$d_{z^2}$ band crosses the Fermi level in DFT+DMFT~\cite{Kitatani2020} and becomes relevant as well. Because of Hund's exchange this orbital cannot be treated as decoupled from the Ni 3$d_{x^2-y^2}$ band. A similar picture has also been observed in other DFT+DMFT calculations~\cite{Karp2020,Pascut2023}.  In $GW$+DMFT the  Ni 3$d_{z^2}$ band touches the Fermi level already at lower dopings at large $k_z$~\cite{Petocchi2020}, and in self-interaction corrected (sic) DFT+DMFT the  Ni 3$d_{z^2}$ band is even more prominent~\cite{Lechermann2019,Kreisel2022}.}. A similar picture has also been advocated in Refs.~\onlinecite{Karp2020}, \onlinecite{Karp2022}, and \onlinecite{Pascut2023}.

\begin{figure}[tb]
    \centering
    \includegraphics[width=0.8\columnwidth]{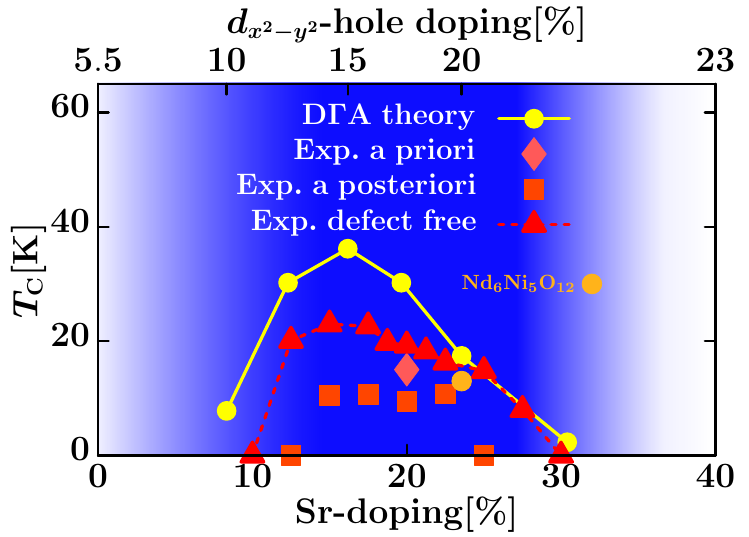}
    \caption{Superconducting phase diagram, $T_c$ vs.~Sr~doping $x$, of Sr$_x$Nd$_{1-x}$NiO$_2$. Following the discovery of nickelate superconductivity (``a priori'', data taken from Ref.~\onlinecite{li2019superconductivity}), $T_c$ was calculated (``D$\Gamma$A theory'', from Ref.~\onlinecite{Kitatani2020}).
      The thus predicted $T_c$ vs.~doping well agrees with the one measured ``a posteriori'' (from Ref.~\onlinecite{Li2020}), especially after ``defect free'' films have been synthesized (from Ref.~\onlinecite{Lee2023}).   Pentalayer  Nd$_6$Ni$_5$O$_{12}$~\cite{pan2021} which has a 20\% doping of the $d_{x^2-y^2}$-orbital~\cite{Worm2021c} also agrees with theory and infinite-layer Sr$_x$Nd$_{1-x}$NiO$_2$ at the same doping of the  Ni  $d_{x^2-y^2}$ orbital (upper $x$-axis). The blue region indicates where only the  Ni  $d_{x^2-y^2}$ orbital and the $A$-pocket cross the Fermi level in multi-orbital DFT+DMFT.}
    \label{fig:PD}
\end{figure}

While the pockets are important for the (Hall) conductivity, we expect superconductivity to primarily emerge from the Ni 3$d_{x^2-y^2}$ band which is strongly correlated.
Indeed, calculations based on this single-band model, with appropriately calculated doping (to account for the pockets),~\cite{Kitatani2020} using the dynamical vertex approximation ($\dga$)~\cite{Toschi2007,Katanin2009,RMPVertex,Kitatani2022} were able to compute the superconducting phase diagram, in good agreement with experiments~\cite{Lee2023}$^,$\footnote{Note that the different substrate LSAT instead of STO~\cite{Lee2023} mainly allows for defect-free nickelate films, as is also obvious from the largely reduced resistivity. The change in lattice parameters is of minor importance, cf.\ the discussion in Section~\ref{sec:rixs_tc}.}, see Fig.~\ref{fig:PD}. In these $\dga$ calculations, antiferromagnetic (AFM) spin fluctuations mediate $d$-wave superconductivity.
Despite the agreement of Fig.~\ref{fig:PD}, 
it is imperative to further test this picture of spin-fluctuation-mediated superconductivity in nickelates. 
An important validation of the spin-fluctuation scenario comes from comparing the spin-wave spectrum predicted by D$\Gamma$A to that measured in experiment. This is the aim of the present paper.

Specifically, signatures of AFM fluctuations have been measured in resonant inelastic X-ray scattering (RIXS)\cite{Lu2021}, nuclear magnetic resonance (NMR)\cite{Cui2021} and
$\mu$SR~\cite{fowlie2022intrinsic} where the fluctuation lifetime can exceed that of the muon~\cite{fowlie2022intrinsic}. Long-range AFM order is, however, absent in infinite-layer nickelates \cite{lin2022universal,Ortiz2021}, a notable difference to cuprates. One natural explanation is the self-doping of the Ni 3$d_{x^2-y^2}$ orbital induced by the $A$- and $\Gamma$-pockets. At hole-doping levels similar to that of the nickelate parent compounds, AFM order has also vanished in cuprates~\cite{Migaku2004,Keimer15}. Consequently, electron doping (or changing a buffer layer to remove the pockets~\cite{Nomura2020,Kitatani2023}) is presumably needed to stabilize AFM order in nickelates. 

In the present paper, we calculate and analytically continue the magnetic susceptibility behind the $\dga$ calculation of Fig.~\ref{fig:PD}.  The non-local scattering amplitude (two-particle vertex) at the heart of 
this magnetic susceptibility directly enters the Cooper (particle-particle) channel as a pairing vertex and thus mediates superconductivity. For details see Ref.~\onlinecite{Kitatani2022}. We also perform RIXS  experiments and compare them to previous RIXS data by  Lu {\it et al.} \cite{Lu2021}. We find theory and experiment to be consistent. In particular, the strength of the experimental AFM coupling is similar to the one extracted from $\dga$, advocating that it is sufficient to mediate the $T_c$ observed in nickelate superconductors. 

The outline of the paper is as follows: 
In Sec.~\ref{sec:theory},
we describe the theory behind the modelling of spin fluctuations and superconductivity by a one-band Hubbard model, the {\it ab-initio} calculated parameters used, and the $\dga$ calculations performed (cf.\ Appendix~\ref{sec:sus}).
Similarly, in  Sec.~\ref{sec:exp}  the experimental methods are discussed, specifically the film growth and RIXS measurements.
Sec.~\ref{sec:caveats} discusses possible shortcomings and sources of errors when extracting the paramagnon dispersion in theory and experiment.
In Sec.~\ref{sec:nickelate_rixs}, we compare the theoretical and experimental
magnetic spectrum. An analysis of the results in terms of a spin-wave model
is presented in Sec.~\ref{sec:spin_wave_picture}. Its interaction dependence 
is elucidated in Sec.~\ref{sec:rixs_tc}, and two possible effects of disorder are discussed in  
in Sec.~\ref{sec:disorder}. Finally, Sec.~\ref{sec:conclusion} summarizes our results.

\section{Methods}
\label{sec:methods}

\subsection{Theory}
\label{sec:theory}
\subsubsection{Modeling}
Previous work, based on DFT+DMFT\cite{Karp2020,Kitatani2020} identified two bands that cross the Fermi-surface in the superconducting regime of infinite-layer nickelates: one band with Ni 3$d_{x^2-y^2}$ character and a pocket around the $A$-momentum composed of Ni 3$d$-$t_{2g}$+Nd-5$d_{xy}$ character (subsequently referred to as the $A$-pocket; for some dopings and rare-earth cations there is also an additional $\Gamma$ pocket). However, the Ni 3$d_{x^2-y^2}$  band and the $A$-pocket within the same cell do not hybridize by symmetry. Hence, to a first approximation, they can be regarded as effectively decoupled. In this picture, superconductivity is expected to emerge from 
the Ni 3$d_{x^2-y^2}$ band, which can be described by a one-band Hubbard model:

\begin{equation}
    \mathcal{H} = \sum_{ij \sigma} t_{ij} \hat{c}_{i\sigma}^{\dagger} \hat{c}^{\phantom{\dagger}}_{j \sigma} + U \sum_{i} \hat{n}_{i \uparrow} \hat{n}_{i \downarrow}.
\label{eq:Hubbard_Model}
\end{equation}
Here, $t_{ij}$ denotes the hopping amplitude from site $j$ to site $i$;
$\hat{c_i}^{\dagger}$ ($\hat{c_j}$) are fermionic creation (annihilation) operators and $\sigma$ marks
the spin; $\hat{n}_{i\sigma} = \hat{c}^{\dagger}_{i\sigma}
\hat{c}^{\phantom{\dagger}}_{i\sigma}$ are occupation number operators. The Coulomb
interaction is, because of screening, restricted to the on-site interaction $U$. The
electrons taken away by the $A$-pocket are accounted for by properly relating the Sr
doping to the $d_{x^2-y^2}$ doping. This translation is displayed in the difference between lower and  upper
$x$-axis of Fig.~\ref{fig:PD}, which is based on multi-orbital DFT+DMFT calculations\cite{Kitatani2020}.

\subsubsection{Ab-initio determined parameters}
\label{sec:parameters}
This Hubbard model has been used sucessfully as an effective low-energy model to calculate the superconducting dome in NdNiO$_2$  \cite{Kitatani2020}.
Here, we  employ exactly the same {\it ab initio}-derived parameters for NdNiO$_2$, see Table~\ref{tab:parameters},
where $t$ is the nearest, $t'$ the next-nearest, and $t''$ the next-next nearest neighbor
hopping amplitude. The tight-binding parameters are obtained after full relaxation of the
lattice parameters with \textsc{VASP}\cite{PhysRevB.48.13115} using the  PBE \cite{Perdew96}
version of the generalized gradient approximation (GGA). In the presence of a substrate, we fix the
in-plane lattice parameters to that of the substrate. 
For this crystal structure, the hopping parameters are subsequently obtained from a DFT
calculation using \textsc{Wien2K} \cite{blaha2001wien2k,Schwarz2002} and
\textsc{Wien2Wannier}\cite{Kunes2010a} to construct maximally localized Wannier
orbitals\cite{Marazari2021}.
As one can see from  Table~\ref{tab:parameters} the variation of these hopping parameters
among different nickelates and substrates is minimal, cf.\ the discussion below. Because of this insensitivity to structural details we restrict ourselves
in the following to one $\dga$ calculation resembling the hopping parameters of  NdNiO$_2$ (bulk).  Specifically, we use the same hopping parameters ($t=0.395\,$eV,  $t'/t=-0.25$, $t''/t=0.12$)  as in Ref.~\onlinecite{Kitatani2020} for calculating the data in Fig.~\ref{fig:PD} and use a temperature $T=60\,$K if not stated otherwise.
  
\begin{table}[tb]
\centering
\begin{tabular}{|c|c|c|c|}
\hline
System & $t$[eV] & $t'/t$ & $t''/t$  \\
\hline
\hline
NdNiO$_2$ (bulk) &0.395 &-0.24  &0.12 \\
\hline
NdNiO$_2$/STO & 0.377  &  -0.25 & 0.13 \\
\hline
NdNiO$_2$/LSAT & 0.392 & -0.25 & 0.13 \\
\hline
LaNiO$_2$ (bulk)  &0.389 &-0.25  &0.12  \\
\hline
LaNiO$_2$/STO  &0.376 &-0.23 &0.11 \\
\hline
LaNiO$_2$/LSAT  &0.390 &-0.22 &0.11 \\
\hline
PrNiO$_2$/STO   & 0.378 & -0.25 & 0.13 \\
\hline
\end{tabular}
\caption{Hopping parameters of the effective Ni 3$d_{x^2-y ^2}$ orbital  for  nickelates
with different spacer cations, two  substrates (STO: SrTiO$_3$; LSAT: (La,Sr)(Al,Ta)O$_3$)
and bulk. }
\label{tab:parameters}
\end{table}

As in Ref.~\onlinecite{Kitatani2020} the one-site Coulomb repulsion $U$ is taken from constrained random phase approximation (cRPA) calculations. A natural first choice would be to simply use $U = U_{\text{cRPA}}(\omega=0)$, which is about $2.6$\,eV for the single-band approximation of LaNiO$_2$\cite{Nomura2019}. However, 
a slightly enhanced static $U$ is needed to empirically compensate for the neglected frequency dependence and increase of $U$ above the effective plasma frequency. Further, previous studies \cite{casula_effmodel,Qiang2018,Honerkamp2018} showed that cRPA overscreens the interaction. To take the above into account and
in agreement with common practice, in  Ref.~\onlinecite{Kitatani2020}
a slightly enhanced value of $U=8$\,$t$ ($3.11$\,eV) was considered as the best approximation. Further, we also use the same scheme to account for the self-doping effect of the $A$-pocket as in Ref.~\onlinecite{Kitatani2020}. That is, the doping of the one-band Hubbard model  is determined from a 5 Ni-$d$ and 5 Nd(La)-$d$ DFT+DMFT calculation for
Sr$_x$Nd(La)$_{1-x}$NiO$_2$, see Supplemental Material of Ref.~\onlinecite{Kitatani2020}.
Both here and in Ref.~\onlinecite{Kitatani2020}, the contribution of the pockets to
superconductivity or the magnetic response, beyond an effective doping, has been neglected.
This is  justified since, due to its strong correlations, the $x^2-y^2$ orbital dominates the magnetic susceptibility and (presumably) the pairing. Also note that the filling of the pockets is low. 

\subsubsection{Magnetic susceptibility}
We compute the magnetic susceptibility $\chi_m$ in $\dga$ for the Hubbard model Eq.~(\ref{eq:Hubbard_Model}) using the parameters motivated in the last section. $\dga$ uses a DMFT solution as a starting point and introduces non-local correlations via the Parquet or, in the simplified version used here, the Bethe-Salpeter equation\cite{Toschi2007,Katanin2009,Valli2015}. We outline in Appendix~\ref{sec:sus}, for the sake of completeness, the steps necessary to obtain $\chi_m$; and refer the reader to Ref.~\onlinecite{RMPVertex} for a more in-depth discussion of the $\dga$,  to Ref.~\onlinecite{Kitatani2022} for details on how to calculate $T_c$, and to Ref.~\onlinecite{Held2018} for a first reading.


On a technical note, we solve the Hubbard model with DMFT using continuous-time quantum Monte-Carlo simulations in the hybridization expansion as implemented in the \textsc{w2dynamics} package \cite{w2dynamics2018}. After DMFT convergence
the two-particle Green's function of the corresponding Anderson impurity model (AIM) is obtained and from it the generalized $\dga$ susceptibility, as outlined in Appendix~\ref{sec:sus}.
To obtain the physical susceptibility we need to sum the generalized susceptibility over two momenta and frequencies and perform the analytical continuation described in the next section.
To ensure good statistics, we use order $10^9$ measurements in a high statistic run for both one and two-particle objects.

\subsubsection{Analytic continuation}
In the $\dga$ calculation, all quantities are defined in terms of imaginary time, or correspondingly, imaginary frequency (Matsubara frequency), see Appendix~\ref{sec:sus}. When comparing to experiments, however, results on the real axis are required. To obtain them we use the open-source package ana$\_$cont \cite{Kaufmann2021}, which employs the maximum entropy (maxent) method  \cite{PhysRevB.44.6011} for bosonic correlation functions like the physical susceptibility in Eq.~(\ref{eq:chi_lambda}). Since the physical susceptibility $ \chi^{q={\mathbf q,\omega}, \lambda}_{m} $ depends on momenta $\mathbf q$ and frequency $\omega$, one analytic continuation of the frequency is performed for each momentum. We fix all hyperparameters of the maxent routine by employing the "chi2kink" method \cite{Kraberger2017} as implemented in ana$\_$cont \cite{Kaufmann2021}.

\subsection{Experiment}
\label{sec:exp}

\subsubsection{Nickelate films}
\label{sec:films}
The precursor films of Pr$_{1-x}$Sr$_x$NiO$_3$ $(x=0,0.2)$ with $8\,$nm thickness were grown on (001)-oriented SrTiO$_3$ and {(LaAlO$_3$)$_{0.3}$(Sr$_2$TaAlO$_6$)$_{0.7}$} {(LSAT) substrates using pulsed laser deposition (PLD). Soft-chemistry reduction using CaH$_2$ powder was then performed to remove the apical oxygens of the precursor films. After reduction ($300^\circ$C, $100$\,min), the perovskite phase was transformed into an infinite-layer phase. The infinite layer Pr$_{1-x}$Sr$_x$NiO$_2$ $(x=0,0.2)$ films were transferred to the PLD chamber to be deposited on their surfaces with $14$-nm thick SrTiO$_3$ protective top layers. The Pr$_{0.8}$Sr$_{0.2}$NiO$_2$ films grown on SrTiO$_3$ substrates are superconducting ones with the onset temperatures of superconducting transition $\sim$$12$~K.}

\subsubsection{RIXS experiments}
\label{sec:RIXS}

Ni $L_3$-edge RIXS measurements were carried out at the I21 beamline at the Diamond Light Source. The energy resolution was set to 39\,meV (full-width-at-half-maximum) at the Ni $L_3$ resonance (850.6\,eV). Incident x-rays with $\pi$ polarisation were used to enhance the paramagnon excitations. The scattering angle was fixed at $154^\circ$ to maximize the in-plane momentum transfer. All RIXS spectra are collected at 16\,K and normalized to the weight of the $dd$ excitations. 

\subsection{Caveats}
\label{sec:caveats}

The $\dga$ calculation of the magnetic susceptibility is, as a matter of course, approximate. The DFT (GGA) starting point puts, e.g., the oxygen orbitals  to close to the Fermi level. This is a bit less relevant for nickelates than for cuprates, and, in particular, when not including the oxygen orbitals in subsequent DMFT calculations (as done here). The next step, DMFT, is restricted to local correlations. Here, DMFT is  used for translating the Sr-doping to the doping of the Ni $3d_{x^2-y^2}$ orbital and for calculating a local vertex of the effective one-band Hubbard model. From this, we then calculate non-local spin fluctuations in $\dga$ through the Bethe-Salpeter ladder, and from these, in turn, the superconducting pairing glue is obtained. This procedure neglects how the superconducting fluctuations in the particle-particle channel feed back to the antiferromagnetic spin fluctuations in the particle-hole channel, and it also presumes that a local frequency-dependent vertex is a reasonably good starting point. Generally, this vertex is much more local than other properties such as the self-energy, even in the superconducting doping regime\cite{RMPVertex}. The good agreement of $\dga$ and diagrammatic quantum Monte-Carlo simulations has been evidenced in Ref.~\onlinecite{Schaefer2021}.

Let us, here, mostly focus on two aspects that we believe are important to keep in mind when comparing the theoretical spectrum of the magnetic susceptibility to RIXS experiments:
(i) The maxent approach is state-of-the-art to solve a {\it per se}  ill-conditioned problem: analytically continuing imaginary time data to real frequencies. Its error grows with frequency, since larger real frequencies only enter exponentially suppressed into the imaginary time (or frequency) data. Further, maxent 
tends to broaden spectra \cite{Kaufmann2021}. Thus, we may expect the theoretical dispersion to be broader and that the maxent error possibly even dominates the widths of the dispersion, especially at higher frequencies.

(ii) On the experimental side, 
RIXS measurements do not probe magnetic excitations exclusively, but rather elementary excitations in general \cite{Ament2011}. To extract the paramagnon dispersion, one fits several functions to the RIXS raw data.
For example, 
the authors of Ref.~\cite{Lu2021} used a Gaussian for the elastic peak, a damped harmonic oscillator (DHO) for the magnon, an anti-symmetrized Lorentzian for phonons and the tail of an anti-symmetrized Lorentzian for the high-energy background, see  supplementary information of Ref.~\onlinecite{Lu2021}. Similarly, {in our fitting function, a DHO convoluted with the energy resolution function is used to model the single magnon. The elastic peak is described with a Gaussian. The smoothly varying background is mimicked by a polynomial function. An additional Gaussian is included to describe the phonon mode at $\sim$$70$~meV when it becomes visible at a large in-plane momentum ($>0.15$~r.l.u.)}. For some dopings and momenta, we have a clear peak structure for the magnon and the error involved in this fitting is mild. In other situations, e.g., close to the $X$ momentum, there is only a minor hump or shoulder, and the magnon energy is much more sensitive to the fitting procedure.

\section{Magnetic response in nickelate superconductors}
\label{sec:nickelate_rixs}

\begin{figure*}[tb]
    \centering
    \includegraphics[width=\textwidth]{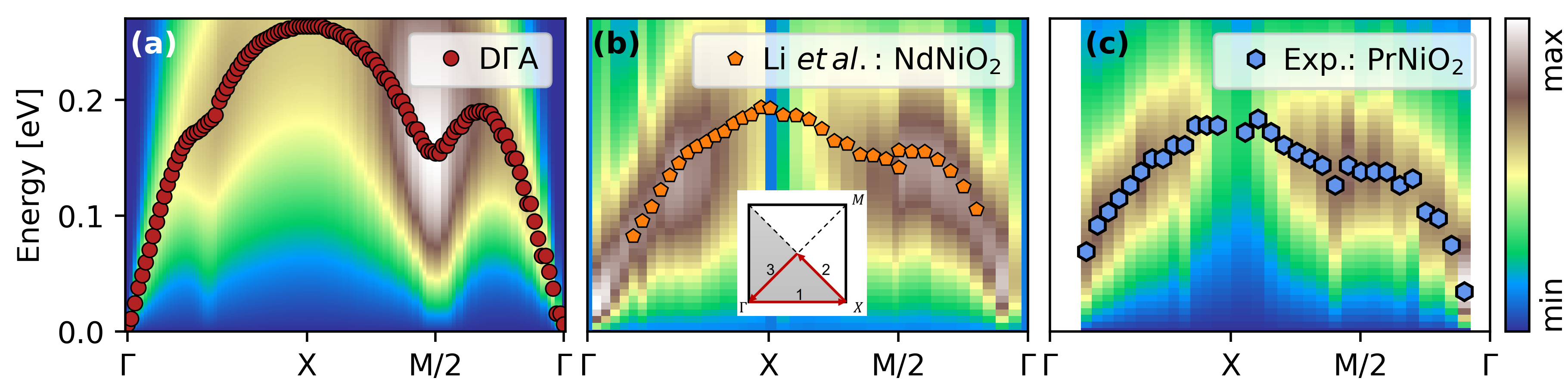}
    \caption{Nickelate magnetic susceptibility (colormap; {arbitrary units}) and paramagnon dispersion (symbols). \textbf{(a)} $\dga$ for NdNiO$_2$ at $T= 60\,$K and $n=0.95$ because of self-doping. \textbf{(b)} RIXS measurements for NdNiO$_2$ on STO at $T = 15$\,K from Ref.~\onlinecite{Lu2021}.
    \textbf{(c)}  Our RIXS measurements for PrNiO$_2$ on STO at $T = {16}$\,K.  Red, orange and blue symbols  mark the peak maximum. Inset in \textbf{(b)}:  Chosen $k$-path through the Brillouin zone, where $\Gamma = (0,0,0)$, $X$ $= (\pi,0,0)$ and $M$/2 $= (\pi/2,\pi/2,0)$. }
    \label{fig:RIXS_x0_Pr}
\end{figure*}

With the good agreement of the theoretical\cite{Kitatani2022} and experimental phase diagrams\cite{Lee2023,Li2020}  in Fig.~\ref{fig:PD}, we here aim at analyzing whether the underlying magnetic fluctuations that mediate $d$-wave superconductivity in theory also agree with experiment. 
The magnetic spectrum and paramagnon dispersion for the two parent compounds  NdNiO$_2$ and  PrNiO$_2$ 
is shown in Fig.~\ref{fig:RIXS_x0_Pr}~(b) and (c), respectively. Here, Fig.~\ref{fig:RIXS_x0_Pr} (a)  displays the imaginary part of the magnetic susceptibility $\chi_{\magn}^{\prime \prime}(\omega,q)$ as computed in $\dga$ using a filling  $n = 0.95$ of the Ni $3d_{x^2-y ^2}$ orbital, originating from the self-doping due to the rare-earth pockets in  NdNiO$_2$.  For  PrNiO$_2$ this self-doping is minimally smaller (3\%). For the hopping parameters and Coulomb interaction, see Sec.~\ref{sec:parameters}.
The dispersion is shown along the high-symmetry path in the Brillouin zone (BZ) from $\Gamma$ to $X$ to $M$/2 to $\Gamma$ that is shown in the inset of Fig.~\ref{fig:RIXS_x0_Pr}(b). The data
of Fig.~\ref{fig:RIXS_x0_Pr}(c) is our own measurement, that of Fig.~\ref{fig:RIXS_x0_Pr}(b) was extracted from the RIXS measurements of
Ref.~\onlinecite{Lu2021}\footnote{As for the dispersion, we plot the fitted damped
harmonic oscillator, which is used as a model to describe the (para)magnon. The fit was
performed by us as described by the authors of Ref.~\onlinecite{Lu2021} in their supplemental
material.}.
Given that we did not adjust any parameters\footnote{{Except for the magnitude, which has been adjusted in both experiments and in theory to a similar scale.}}, the agreement between theory and the
magnon dispersion extracted from RIXS is quite good. This indicates that experimental spin
fluctuations are similar to those leading to $d$-wave superconductivity in the $\dga$
calculations.

Looking more into the details, we see that the overall paramagnon bandwidth is systematically a bit larger in $\dga$. For example, the peak
of $\chi_{\magn}(q=X,\omega)$ is at $\omega_{\rm peak}\sim 260$\,meV in $\dga$, while the
measured one is close to $\sim 190$\,meV. While the tendency of the maximum entropy method
to broaden spectra might also slightly affect the position of the maximum and the spectrum
around the $X$ point is more blurred in theory and experiment, overall the difference is
beyond the maxent error. In agreement with the overall width, the slope of the linear dispersion around $\Gamma$ deviates somewhat and
hence we conclude: the overall width of the paramagnon dispersion in theory is noticeable larger than in
experiment.
This difference corresponds to a larger  effective spin coupling $J$ in  $\dga$, as we discuss in more detail  in Sec.~\ref{sec:spin_wave_picture}. 

Furthermore, the ``dip'' observed in the dispersion around the $M$/2 momentum, which corresponds to
a next-next nearest neighbor exchange $J''$ in a spin-wave picture, is more pronounced in theory than in experiment.  
In Sec.~\ref{sec:rixs_tc}, we show that using a larger $U=9\,t$ (instead of $U=8\,t$) results in a better agreement of the spin wave dispersion and also of the
phase diagram of Sr$_x$Nd$_{1-x}$NiO$_2$ on STO, which has considerably lower $T_c$'s than  Sr$_x$Nd$_{1-x}$NiO$_2$ on LSAT. Please note that the origin for this experimental difference is not the minute change in lattice constant, but that growing  Sr$_x$Nd$_{1-x}$NiO$_2$ on LSAT results in cleaner films without stacking faults \cite{Lee2023}. These ``defect-free'' films have a much lower resistivity and higher $T_c$'s, and agree better with our best estimate $U=8\,t$.

\begin{figure}[tb]
    \centering
    \includegraphics[width=\columnwidth]{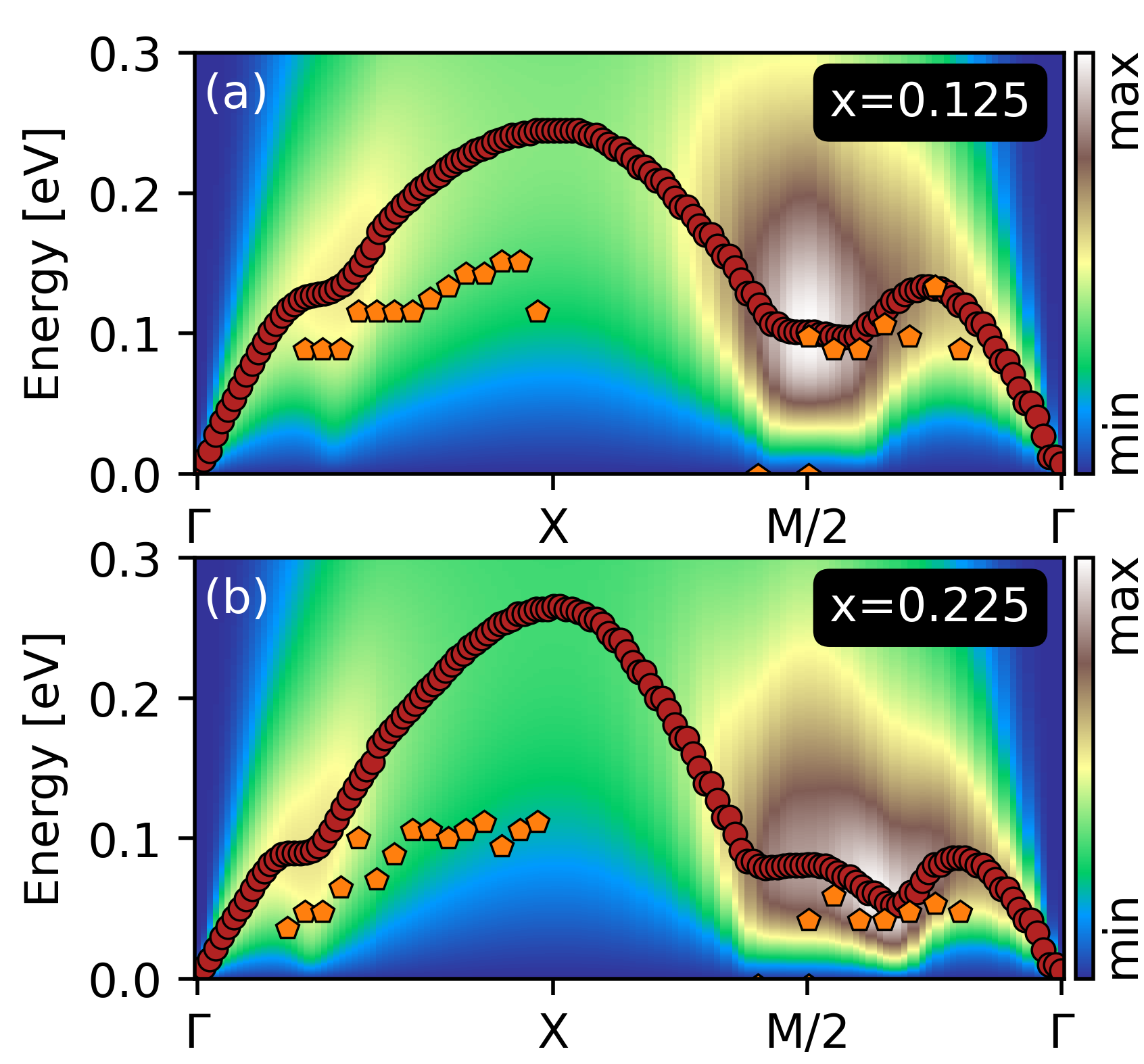}
    \caption{Paramagnon dispersion of Sr$_x$Nd$_{1-x}$NiO$_2$ for  \textbf{(a)} $x = 0.125$ ($n=0.875$)  and  \textbf{(b)} $x \simeq 0.225$ ($n=0.8$) in $\dga$. The red dots mark the maximum of the $\dga$ dispersion at each $k$-point, while the orange pentagons are the corresponding RIXS peak locations from Ref.~\onlinecite{Lu2021}.}
    \label{fig:RIXS_x0_x0125_science}
\end{figure}

Let us now turn toward the doped compounds. $\dga$ results for Sr$_{0.2}$Nd$_{0.8}$NiO$_2$
 with $x = 0.125$ (effective filling $n = 0.875$)
 and $x = 0.225$ ($n = 0.80$) are displayed in Fig.~\ref{fig:RIXS_x0_x0125_science}(a) and (b), 
 respectively, and compared to the experimental dispersion (orange pentagons). Consistent with experiment, we observe a shift towards lower energies around the $M$/2 momentum. Furthermore, the 
 amplitude of $\chi_{\magn}$ decreases as $q \to$$X$, which was also observed in Ref.~\onlinecite{Lu2021}\footnote{See supplementary material of Ref.~\onlinecite{Lu2021} Fig.~S6(b).}.
%
%

Similar to the parent compound (Fig.~\ref{fig:RIXS_x0_Pr}), the bandwidth  in $\dga$ is larger also at finite doping. Particularly the peak position at the $X$ momentum shows a substantial deviation compared to the one extracted from RIXS. This may be partially (but not fully) attributed to the bias introduced both on the theoretical and experimental sides. 
On the one hand, we expect a worse performance of the numerical analytic continuation for large frequencies. A spectrum already relatively flat at the $X$ momentum might be additionally broadened because of the maximum entropy analytic continuation. On the other hand, the intensity of the paramagnon peak is also reduced in RIXS \cite{Lu2021}, making the experimental fitting procedure more difficult.
Notwithstanding, there is  a larger theoretical dispersion (or $J$) than in experiment. Qualitatively similar, $\dga$ and RIXS show that the minimum at the {$M$/2}  momentum of the parent compounds turns into a flat dispersion or even a local maximum with doping.

\section{Discussion}
\label{sec:discussion}

\subsection{Effective spin-wave picture}
\label{sec:spin_wave_picture}

In the limit of a large Hubbard interaction $U$, the Hubbard model (Eq.~\ref{eq:Hubbard_Model}) reduces to an effective Heisenberg Hamiltonian. 
We refer the reader to  Ref.~\onlinecite{Delannoy2009} and references therein for an extensive discussion for the one-orbital Hubbard model.
While this mapping provides a direct relation between $t$, $U$ and the effective spin couplings $J$, the temperature does not enter, nor does the doping. Indeed, strictly speaking, the mapping onto the spin model is possible only for an insulator (at half-filling). Yet, the parent compound Nd(Pr,La)NiO$_2$ is --in contrast to cuprates-- neither half-filled nor  Mott-insulating  because of the finite pockets. Furthermore, the  mapping onto a spin model becomes rather tedious in the presence of hoppings $t'$ and $t''$ beyond nearest-neighbors \cite{Delannoy2009}. 
Nevertheless, the spin model and the spin-wave dispersion  provide a somewhat intuitive picture for understanding the characteristics of spin fluctuations also in the present case of nickelates.

For these reasons, we employ here the same approach as in experiment also for the $\dga$ data: that is, we fit the  spin-wave dispersion of the
Heisenberg model to our $\dga$ results in order to extract information about effective spin-couplings $J$ and $J'$.
Including only the nearest neighbor ($J$) and next-nearest neighbor spin-exchange
$J'$, the effective classical spin-wave dispersion for a spin-$1/2$ system is given by
\cite{Coldea2001,Delannoy2009,Ivashko2019}
\begin{equation}
\omega_{\textbf{k}} = Z_C \sqrt{\textrm{A}_{\bk}^2 - \textrm{B}_{\bk}^2},
\label{eq:spin_excitation_dispersion}
\end{equation}
where $Z_C$ is the spin-wave renormalization factor that accounts for the effects of quantum fluctuations and 
\begin{equation}
\begin{split}
    \textrm{A}_{\bk} &= 2 J + 2 J' [\cos(k_x)\cos(k_y) -1], \\
    \textrm{B}_{\bk} &= J [\cos(k_x) + \cos(k_y)].
\end{split}
\label{eq:dispersion_terms}
\end{equation}

\begin{figure}[tb]
    \centering
    \includegraphics[width=\columnwidth]{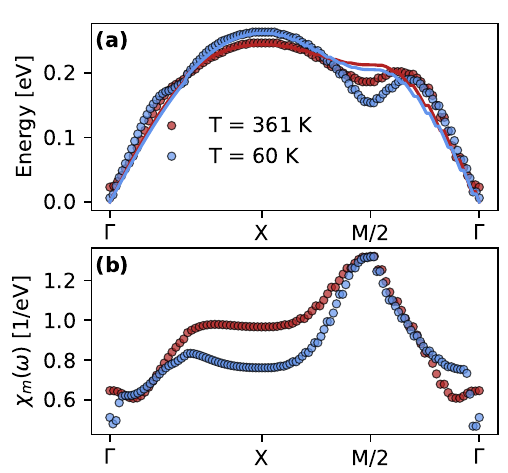}
    \caption{Paramagnon dispersion obtained in $\dga$ for LaNiO$_2$ as a function of temperature. \textbf{(a)} $\dga$ dispersion (dots) and spin-wave fit (lines) using Eq.~(\ref{eq:spin_excitation_dispersion}). \textbf{(b)} Magnetic susceptibility $\chi_m(\omega_{\textrm{peak}})$ at the peak frequency $\omega_{\textrm{peak}}$ of panel \textbf{(a)}. Red:  $361$\,K; blue:  $60$\,K.}
    \label{fig:magnon_temperature_evolution}
\end{figure}

To better compare with the values obtained in experiment, we fix $Z_C = 1.18$ as in Ref.~\onlinecite{Lu2021}. Fig.~\ref{fig:magnon_temperature_evolution} shows the paramagnon dispersion and the corresponding value of the magnetic susceptibility $\chi_m(\omega_{\textrm{peak}})$.

To a first approximation (if $J'\ll J$), the width of the spin wave dispersion is $2J$ in Eq.~(\ref{eq:spin_excitation_dispersion}), with $\omega_{{\mathrm peak}\,{\mathbf k}}=\omega_{{\mathrm peak}\,M/2}=2J$.  A finite $J'$ adds a skewness to this as  $\omega_{{\mathrm peak} \, X}=2J-4J'$, whereas   $\omega_{{\mathrm peak}\,M/2}=2J-2J'$. The fact that the maximum of the dispersion is at ${\mathbf k}=X$ thus implies a ferromagnetic (negative) $J'$. This is qualitatively similar to cuprates which have, however, a considerably  larger $J$  (e.g., $J=112\,$meV and $J'=-15\,$meV for La$_2$CuO$_4$\cite{Coldea2001}). It is, on the other hand, different from other nickelates such as La$_2$NiO$_4$\cite{XXXIzabelas_paperXXX} that show an antiferromagnetic (positive) $J'$ and opposite skewness. This is because the latter require a multi-band description with Ni 3$d_{x ^2-y ^2}$ and Ni 3$d_{z^2}$ orbital, resulting in a larger effective $U$\cite{XXXIzabelas_paperXXX}.

The $\dga$ skewness is well described by the spin-wave fit, but there is  a pronounced minimum at the $M$/2-point  (for the parent compound) which is not well captured by the ($J$-$J'$) spin-wave fit. We presume higher-order couplings, which are difficult to fit to the numerical $\dga$ data,  are needed to account for 
such a minimum. Specifically, a next-next neighbor exchange $J''$ adds a term
\begin{equation}
  \textrm{A}_{\bk} \rightarrow  \textrm{A}_{\bk} -2J'' (1-[\cos(2k_x)+\cos(2k_y)]/2)
\end{equation}
in  Eq.~(\ref{eq:spin_excitation_dispersion})
\cite{Coldea2001}.
For positive (antiferromagnetic) $J''$ this results in the observed  local minimum at $M/2=(\pi/2,\pi/2)$. The change of this minimum to a maximum as observed with doping, then implies 
a change of sign of $J''$.

Similar to the cuprate superconductor La$_2$CuO$_4$ \cite{Coldea2001} we observe that the dispersion along the antiferromagnetic zone boundary becomes more pronounced as temperature is lowered. This mode-hardening is mimicked by a {\it ferromagnetic} next-nearest spin-coupling whose strength increases from $J'= -13$\,meV at $361$\,K to $J' = -23$\,meV at $60$\,K. For $J$ the trend is opposite and its value gets reduced from $76$\,meV at $361$\,K to $62$\,meV at $60$\,K. Experimentally, $J = 63.6 \pm 3.3$\,meV and $J' = -10.3 \pm 2.3$\,meV at $20$\,K were reported in Ref.~\onlinecite{Lu2021} for NdNiO$_2$ . Similarly, we obtain {$J = 70.0 \pm 5.5$\,meV and $J' = -8.0 \pm 3.8$\,meV at $16$\,K} from RIXS for PrNiO$_2$, see Table~\ref{tab:spin_couplings_parent}.

\begin{table}[]
\centering
\begin{tabular}{|c|c|c|c|c|c|c|}
\hline
$U $ [t] & $8$ & $8$ & $9$ & $10$ &  RIXS PrNiO$_2$ &  RIXS NdNiO$_2$  \cite{Lu2021} \\
\hline
$T $ [K] & $361$ & $60$ & $60$ & $60$ & {16} & $20$\\
\hline
\hline
$J$ [meV]          & $76$    & $62$   & $64$   & $44$    & {$70.0 \pm 5.5$}   & $63.6 \pm 3.3$ \\
\hline
$J'$ [meV]         & $-13$   & $-23$ & $-12$   & $-12$    & {$-8.0 \pm 3.8$}   & $-10.3 \pm 2.3$ \\
\hline
\end{tabular}
\caption{Effective spin-coupling $J$ and $J'$ for NdNiO$_2$ {and PrNiO$_2$} obtained with $\dga$ and
measured in RIXS \cite{Lu2021} given in units of meV. We list results for fits to $\dga$
dispersions at different interaction values $U =\{8,9,10\}$ in units of the hopping $t =
0.389$\,eV and two different temperatures $T = \{300\,\text{K}, 60\, \text{K}\}$ for
$U=8$\,$t$.}
\label{tab:spin_couplings_parent}
\end{table}


\subsection{Interaction dependence}
\label{sec:rixs_tc}
Table~\ref{tab:spin_couplings_parent} suggests that the $\dga$-fitted $J$ value
agrees better with experiment if a larger $U=9\,t$ value is considered. Indeed, $U=9\,t$  has been
considered in Ref.~\onlinecite{Kitatani2020} to be the maximal $U$ still consistent with
the cRPA,  while $U=8\,t$ is the best estimate. 

\begin{figure*}[tb]
\begin{minipage}{.68\textwidth}
  \centering
    \includegraphics[width=\columnwidth]{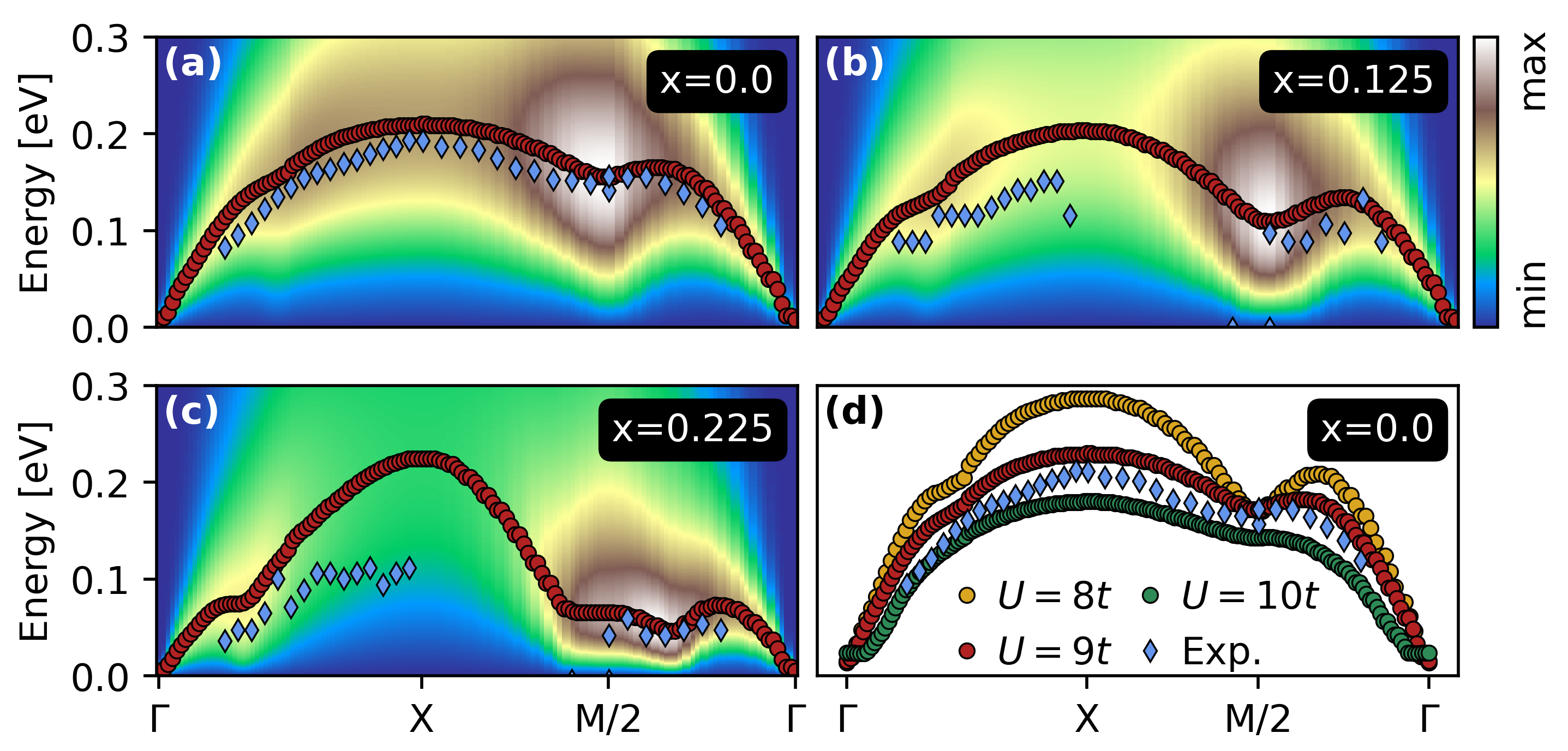}
\end{minipage}
\hfill
\begin{minipage}{.31\textwidth}
  \caption{Paramagnon dispersion of Sr$_x$Nd$_{1-x}$NiO$_2$ using a larger interaction $U = 9\,t \simeq 3.5$\,eV. \textbf{(a)} Parent compound $x = 0.0$ ($n = 0.95$) at $T \simeq 60K$; \textbf{(b)} $x = 0.125$ ($n=0.875$); \textbf{(c)} $x = 0.225$ ($n=0.8$). The red dots mark the maximum of the dispersion at each $k$-point, while the blue dots are the corresponding peak locations measured in RIXS (taken from Ref.~\onlinecite{Lu2021}).  \textbf{(d)} Paramagnon dispersion of the parent compound for different values of the interaction $U$. }
    \label{fig:RIXS_science_U9}
\end{minipage}
\end{figure*}

 Fig.~\ref{fig:RIXS_science_U9} shows the magnetic susceptibility calculated in the same way as in Fig.~\ref{fig:RIXS_x0_Pr} and Fig.~\ref{fig:RIXS_x0_x0125_science}, but now for an interaction value $U = 9\,t$. The colormaps in (a,b,c) show $\chi_{\magn}^{q,\omega}$ along the same high-symmetry path as shown in the inset of Fig.~\ref{fig:RIXS_x0_Pr}(b). The red dots mark the maximum at each momentum, while the blue diamonds correspond to the peak maxima reported in RIXS~\cite{Lu2021}. Finally, Fig.~\ref{fig:RIXS_science_U9}(d) compares the peak location of the paramagnon dispersion for several interaction values.

\begin{figure}
    \centering
    \includegraphics[width=\columnwidth]{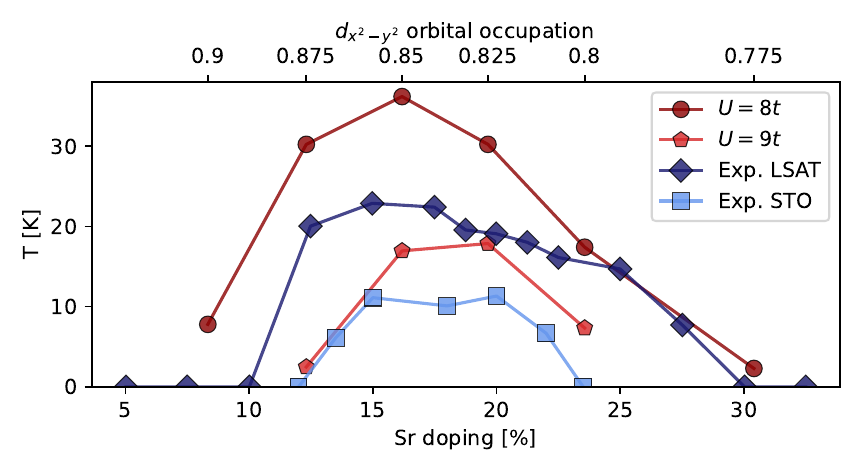}
    \caption{Superconducting phase diagram $T_c$ as a function of Sr doping for Sr$_x$Nd$_{1-x}$NiO$_2$, comparing  $\dga$ with $U=8\,t$($U=9\,t$) from Ref.~\onlinecite{Kitatani2020} to
 experiment on a LSAT(STO) substrate from Ref.~\onlinecite{Lee2023} (Ref.~\onlinecite{Li2020}). The increase of $T_c$ in $\dga$ when using LSAT instead of STO in-plane lattice parameters is much weaker; the difference between the two substrates hence most likely reflects the improvement thanks to cleaner, defect-free films for the LSAT substrate.}
    \label{fig:phase_diagram_u8_U9_comparison}
\end{figure}

 The overall width of the dispersion is reduced for larger $U$, as expected from the spin-wave picture discussed in the previous subsection. Subsequently, the agreement with experimental measurements is improved compared to the results of $U = 8$\,$t$ in Fig.~\ref{fig:RIXS_x0_Pr} and Fig.~\ref{fig:RIXS_x0_x0125_science}.
 This seemingly suggests that $U = 9$\,$t$ is more appropriate (but cf.\ our discussion below). 
Let us, also, compare the superconducting transition temperature and discuss it along with the paramagnon dispersion. Fig.~\ref{fig:phase_diagram_u8_U9_comparison} shows the $\dga$ phase diagram for the two interaction values $U = 8\,t$ (light red) and $U = 9\,t$ (dark red)  together with two experimentally measured domes (blue curves). The theoretical values are taken from Ref.~\onlinecite{Kitatani2020} and the experimental ones are from Ref.~\onlinecite{Lee2023} [NdNiO$_2$ on  LSAT]  and  Ref.~\onlinecite{Li2020} [NdNiO$_2$ on STO].

The measurements for the same nickelate NdNiO$_2$ on different substrates, one  on STO
\cite{Li2020} (light blue) and one on  LSAT \cite{Lee2023} (dark blue), show about a
factor of two difference in the superconducting $T_c$. The higher $T_c$  for the LSAT
substrate is, by the authors of Ref.~\onlinecite{Lee2023},  attributed to the difference in
film cleanliness, with fewer lattice defects observed for the LSAT substrate. This conclusion is supported by scanning
tunneling microscopy that show fewer Ruddlesden-Popper stacking faults. Further support comes from a 
strongly reduced resistivity for the samples grown on LSAT. A dependence of
$T_c$ on the residual resistivity, which was taken as a proxy measure for disorder and
lattice defects, was also reported in Ref.~\onlinecite{Hsu2022} and cleaner films show a
larger $T_c$ in a manner ``not too different from cuprate superconductors''. Indeed,
cuprates show a decrease in $T_c$ for an increasing in-plane \cite{Fukuzumi1996},
out-of-plane  \cite{Hobou2009,Eisaki2004,Fujita2005} resistivity and when magnetic or
non-magnetic impurities are introduced \cite{Bonn1994}.

On the theoretical side, we observe a similar difference in transition temperature between
calculations using $U = 8$\,$t$ and $U = 9$\,$t$, respectively. However, the respective
Wannier tight-binding parameters for LaNiO$_2$\footnote{It is common practice to use La
instead of Nd to avoid the (wrong) appearance of $f$-bands around the Fermi energy. For
treating Nd directly the Nd $f$-shell electrons are usually considered as core electrons.}
with in-plane lattice constants fixed to those of STO and LSAT are quite similar, see
Sec.~\ref{sec:parameters}. That is, we find that the nearest-neighbor hopping $t$
increases by about $\sim 4\%$ from STO to LSAT, while the ratios of $t'/t$ and $t''/t$
remain essentially the same. An increase of $t$ is not surprising as the smaller in-plane
lattice constant of LSAT increases the orbital overlap. 
On the other hand, the Hubbard interaction ($U_{\text{cRPA}} = 2.6$\,eV) essentially does not change
when performing constrained random phase approximation calculations for LaNiO$_2$ with the
$a-b$ lattice parameters fixed to either that corresponding to LSAT or that of STO, respectively. 

Considering these changes of the effective single-band Hamiltonian, we expect samples grown on LSAT to have an intrinsically larger $T_c$, since $t$ sets the energy scale and a smaller $U/t$ is also beneficial \cite{Kitatani2020}. That being said, the expected difference in $T_c$, as a result of the slightly different intrinsic models, is closer to $10-15\%$\footnote{About $\sim 4\,\%$ because of a change in $t$ and another $5-10\,\%$ because of the change in $U/t$.}, but almost certainly not a factor of two \footnote{Larger changes of the hopping parameters are possible by applying pressure, see Ref.~\onlinecite{DiCataldo2023}.}.  For this reason we conclude that changes in our effective single-band Hubbard model do not explain  differences in the measured $T_c$'s for different substrates. The difference has to lie somewhere else, and the reduced number of defects when growing nickelates on  LSAT is the most likely explanation for the enhanced $T_c$ and reduced resistivity, as also originally suggested in Ref.~\onlinecite{Lee2023}.

Following this argument, the appropriate Hubbard interaction for the effective single-band description of infinite-layer nickelates should be close to  our best estimate $U = 8\,t$  with an intrinsic $T_c^{\text{max}} \simeq 30$\,K, comparable to that measured on LSAT\footnote{Effects beyond the single-band model will always lead to some discrepancy. For this reason, we refrain from fine-tuning parameters.}. Consequently, we would expect the $T_c$ of samples grown on STO to be similar (within 4\%) once sample of comparable quality are synthesized. What remains to be understood is how defects and lattice disorder influence the paramagnon dispersion and $T_c$.

\subsection{Effect of disorder and stacking faults on the paramagnon dispersions}
\label{sec:disorder}

To fully address the influence of impurities and lattice defects on the paramagnon dispersion, large-scale calculations for supercells that include these defects would be required. Such calculations are not feasible at the moment, at least not for $\dga$ calculations or similar many-body methods that include non-local fluctuations. For this reason, we will restrict ourselves here instead to qualitative considerations. 

One possibility is that defects reduce the effective antiferromagnetic coupling strength $J$ and, with reduced antiferromagnetic fluctuations, also $T_c$. While estimating the absolute influence of such local defects in RIXS is very difficult, if not impossible, samples that show a different $T_c$ can be compared. Such a study would include measurements of several samples of the same ``species'', e.g., Sr$_{0.2}$Nd$_{0.8}$NiO$_2$ on STO, which show a sample-to-sample variation in $T_c$. Along the same lines, comparing the paramagnon dispersions for samples grown on different substrates (e.g., STO and LSAT) would yield valuable information about the connection between the paramagnon dispersion and $T_c$. A study similar to the latter has already been performed for the related PrNiO$_2$ compound \cite{Qiang2022}. The measurements suggest that the paramagnon dispersion and $J$ are similar for samples grown on LSAT and STO. However, those measurements were done on the non-superconducting parent compound. Hence, it would be interesting to check if the reported results remain unchanged if samples with different $T_c$'s are measured directly. Let us, in this context, mention that  for cuprates it was possible to correlate the increase of $T_c$ with the increase of $J$ for different cuprates\cite{WangCuprates2022}. Along this line of thinking, the better agreement of $T_c$ and the RIXS spectrum of  NdNiO$_2$ on STO for the larger $U=9\,t$  might just mimic the suppression induced by  disorder which is not included in our calculations.

\begin{figure*}[tb]
  \begin{minipage}{.68\textwidth}
    \centering
    \includegraphics[width=0.95\columnwidth]{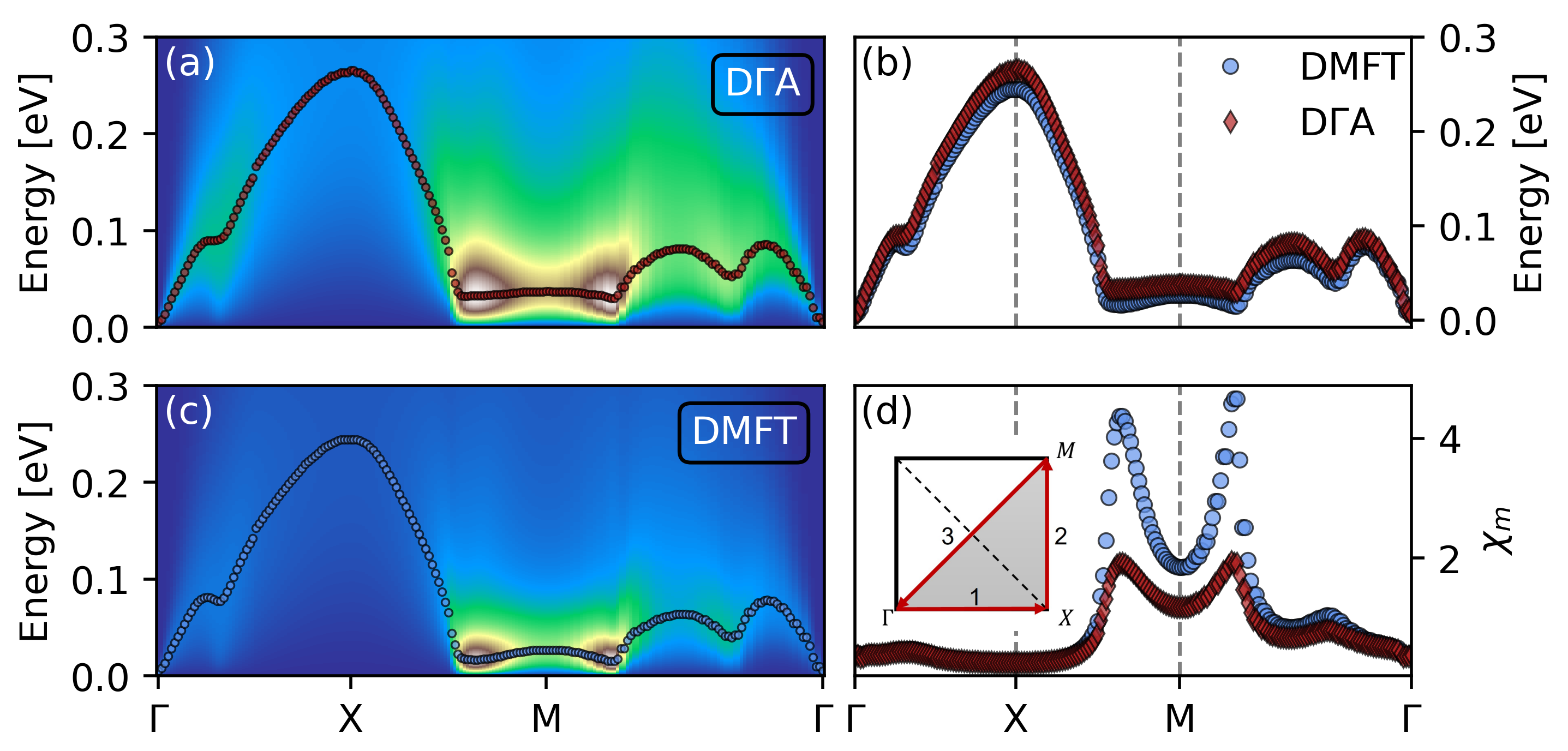}
  \end{minipage}\hfill
  \begin{minipage}{.31\textwidth}
    \caption{Paramagnon dispersion of Sr$_x$Nd$_{1-x}$NiO$_2$ for $x \sim 0.225$  ($n=0.8$) within the effective single-orbital Hubbard model scenario. \textbf{(a)} Colormap of $\chi_{\magn}^{q,\omega}$ calculated by \textbf{(a)} $\dga$ and \textbf{(c)}  DMFT. \textbf{(b)} Comparison of the maximum of the dispersion corresponding to the lines on top of the colormaps in  \textbf{(a)} and \textbf{(c)}, respectively. \textbf{(d)} Peak value of the magnetic susceptibility;  inset: $k$-path in the BZ, where $\Gamma = (0,0,0)$, $X$ $= (\pi,0,0)$ and $M$ $= (\pi,\pi,0)$.
      The main difference between DMFT and $\dga$ is the reduced correlation length.  Parameters: $t = 0.389\,$eV, $t'/t=-0.25$, $t''/t = 0.12$, $U = 8\,t$, $T=60\,$K.}
    \label{fig:RIXS_n080}
  \end{minipage}
\end{figure*}

Let us also point out 
another way how lattice defects might influence $T_c$: decreasing the magnetic correlation length $\xi$. Particularly, if  stacking faults and similar defects introduce artificial ``grain boundaries''\footnote{See for examples Fig.~1 in Ref.~\onlinecite{Lee2023}.}, $\xi$ might be restricted to stay below the typical grain-boundary distance,
without directly changing the effective antiferromagnetic coupling strength $J$.
Though conceptually somewhat different, the $\lambda$-correction in $\dga$ \cite{Katanin2009} has a similar effect in the sense that $\lambda$ causes a decrease in the magnetic correlation length. 
Such a reduced correlation length (added mass), however, essentially does not change the paramagnon dispersion, see Fig.~\ref{fig:RIXS_n080} and the discussion in the next paragraphs. Furthermore, the intensity primarily changes around the M $=(\pi,\pi)$ momentum, where the susceptibility and $\xi$ are the largest.

Such an effective paramagnon-mass enhancement or reduced $\xi$ is difficult to extract
from RIXS, which cannot access the $M$ point in nickelates. Yet, it is precisely the strength of the
susceptibility around the $M$ momentum, which,  from a spin-fluctuation or $\dga$
perspective, is most important for $T_c$. The $M$ point and a prospective difference in
correlation length $\xi$ for different substrates might be accessible in neutron
scattering.
If measurements of the magnetic correlation length $\xi$ for superconducting samples grown on different substrates show a suppressed $\xi$ at the M point for samples grown on STO compared to those grown on LSAT, this would support this second disorder scenario. 

To investigate the influence of a suppressed correlation length onto the paramagnon dispersion, we compare $\chi_m$ between DMFT  and $\dga$  in Fig.~\ref{fig:RIXS_n080} along a high-symmetry path through the BZ now including the $M$ momentum [see inset in panel (d) for the Brillouin zone]. We choose the overdoped compound ($x=0.225$) since DMFT shows no antiferromagnetic order for this doping. Hence, we can directly compare $\chi_{\magn}$ between DMFT  (no $\lambda$ correction) and $\dga$ (with $\lambda$ correction), cf.\ Ref.~\onlinecite{RMPVertex}.  The DMFT with a large correlation length $\xi$ is shown in panel (c), while $\dga$ with a shorter $\xi$ is displayed the latter in panel (a).

We draw the location of the peak in $\chi_{\magn}$ at each momentum in Fig.~\ref{fig:RIXS_n080}(b). It remains essentially unchanged in the presence of a $\dga$ $\lambda$-correction, i.e., with a reduced 
$\xi$. The magnitude of the respective susceptibility at the peak is, however, drastically different around the M momentum, as can be seen in Fig.~\ref{fig:RIXS_n080}(d). A suppression of $\chi_m$ is strongest when the DMFT susceptibility is large, which is no surprise since $\chi_m \rightarrow  1/(1/\chi_m +  \lambda)$ in $\dga$. Hence, a reduction of the correlation length for antiferromagnetic fluctuations is expected to be virtually invisible in RIXS measurements, which do not access the M point. At the same time, the reduction of $T_c$ can be dramatic.

Let us further note that we observe incommensurate antiferromagnetic fluctuations, evidenced by the shift of the maximum amplitude slightly away from $M$ in Fig.~\ref{fig:RIXS_n080}(d). This is typical for overdoped cuprates \cite{Lipscombe2009,Fujita2012}$^,$\footnote{Let us note that the prototypical cuprate L$_2$CuO$_4$ is a 214 layered system compared to the 112 layer structure of nickelates.}. For nickelates, measurements that could distinguish commensurate from incommensurate spin fluctuations have, to the best of our knowledge, not been performed yet.

\section{Conclusion}
\label{sec:conclusion}

The pairing symmetry obtained in $\dga$ for nickelates is $d$-wave, reminiscent of that in cuprates \cite{Keimer15}.
   On the experimental side, the pairing symmetry of nickelates remains an open question as results are still inconclusive\cite{Gu2020b,Harvey2022,Chow2022}.
Also the mechanism for superconductivity remains highly controversial, not only for  nickelates but also for cuprates. Many different mechanisms have been proposed \cite{Lee2006,Scalapino12,Fradkin2015,Keimer15}. In $\dga$ spin fluctuations mediate superconductivity\footnote{
  Cf.\ Refs.~\onlinecite{Kitatani2019} and \onlinecite{Dong2022}
  for an in-depth analysis of the pairing vertex in the Hubbard model and its  fluctuation diagnostic \cite{Schaefer21-2}.}.
%
%

Further, in the case of nickelates, even the minimal model is hotly debated.
Among others, the relevance of multi-orbital physics \cite{Kreisel2022,Werner2019,Petocchi2020}, Kondo physics \cite{Zhang2022}, or even phonons \cite{Alvarez2022} have been suggested.
The single-band Hubbard model with an appropriately calculated doping of the Ni 3$d_{x^2-y
^2}$ orbital, that was used in the present paper, is arguably the simplest possible model
for spin fluctuations and superconductivity in nickelates. It correctly reproduces the doping range
of superconductivity, the absolute value of $T_c$, and also the skewness of the phase
diagram. Here, the skewness is a consequence of the largely decoupled
ligand pockets which accommodate part of the holes from the Sr-doping in a non-linear fashion. This skewness of the superconducting dome is a notable
difference to cuprates.

Given the good agreement of the superconducting phase diagram and 
antiferromagnetic  fluctuations at its origin, a critical check whether or not spin fluctuations with similar characteristics are observed in experiment is imperative. To achieve this validation, we compared the paramagnon dispersion calculated in $\dga$ with the one extracted from RIXS, both from our measurements and those from Ref.~\onlinecite{Lu2021}. We find the spin spectrum is overall similar, especially when considering biases  expected both from theory and experiment. This means that the experimental
spin fluctuations are consistent with the observed $T_c$ in nickelates within the spin-fluctuation scenario of superconductivity. Then, the weaker  spin
fluctuations (or $J$) in nickelates, compared to cuprates, might also explain their lower $T_c$.

The agreement is however not perfect. The total width of the paramagnon dispersion (or $J$) is somewhat smaller in experiment than in theory. Using an enhanced Coulomb interaction $U=9\,t$ in $\dga$ much better matches the RIXS  spectrum and also the superconducting phase diagram of Sr$_x$Pr$_{1-x}$NiO$_2$ and Sr$_x$Nd$_{1-x}$NiO$_2$ on STO. However, this might be by chance. The larger $T_c$'s are observed for Sr$_x$Nd$_{1-x}$NiO$_2$ on a LSAT substrate. The difference of  LSAT and STO in-plane lattice constants  and thus also the $\dga$ hopping parameters are way too small to explain the by a factor-of-two higher $T_c$.

Having much less defects  (Ruddlesden-Popper stacking faults\cite{Lee2023})
is instead expected to be the origin of the higher $T_c$ and better conductivity
of Sr$_x$Nd$_{1-x}$NiO$_2$ on LSAT. Increasing $U$ might only be an imperfect way to mimic
this disorder reduction of antiferromagnetic spin fluctuations and $T_c$. As discussed in
Sec.~\ref{sec:disorder}  local disorder can reduce $J$. From this perspective,
Sr$_x$Nd$_{1-x}$NiO$_2$ on  LSAT with less disorder should show a larger $J$ and
dispersion in RIXS. However, if the disorder is rather cutting-off the spin correlation
length --and Ruddlesden-Popper stacking faults might hint in this direction--, we show
that the effect on the RIXS dispersion can be negligible. In this scenario, disorder only
affects the peak height at $M=(\pi,\pi)$, which is not accessible in RIXS
but in neutron scattering experiments.

In all, we believe that our joint theoretical and experimental investigation strengthens the case for spin-fluctuation mediated superconductivity in nickelates.

{\it Acknowledgments}
We thank Simone Di Cataldo,   Oleg Janson, and Jan Kune\v{s} for helpful discussions. We further acknowledge funding through the Austrian Science Funds
(FWF) projects I 5398, I 6142, P 36213, SFB Q-M\&S (FWF project ID F86), {the Swiss National Science Foundation under 200021$\_$188564}, Grant-in-Aids for Scientific Research (JSPS KAKENHI) Grants No. JP21K13887 and No. JP23H03817, {the Research Grants Council of Hong Kong (ECS No. 24306223), and Research Unit
QUAST by the
Deutsche Foschungsgemeinschaft (DFG; project ID FOR5249) and FWF
(project ID I 5868). L.S.\ is thankful for the starting funds from
Northwest University. {I.B.\ acknowledges support from the Swiss Government Excellence Scholarship under the project number ESKAS-Nr:~2022.0001.} Calculations have been done in part on the Vienna
Scientific Cluster (VSC).
For the purpose of open access, the authors have applied a CC BY public copyright licence to any Author Accepted Manuscript version arising from this submission.

\appendix
\section{Magnetic susceptibility in $\dga$}
\label{sec:sus}
As a first step to calculate the magnetic susceptibility $\chi_m$,
we solve the Hubbard model Eq.~(\ref{eq:Hubbard_Model}) within DMFT and subsequently sample the local two-particle Green's function ($G^{(2)}$) for the corresponding Anderson impurity model (AIM): 
\begin{widetext}
    \begin{equation}
        G^{2}(\nu_1,\nu_2,\nu_3,\nu_4) = \int_{0}^{\beta} \int_{0}^{\beta} \int_{0}^{\beta} \int_{0}^{\beta} d\tau_1 d\tau_2 d\tau_3 d\tau_4 G^{2}(\tau_1,\tau_2,\tau_3,\tau_4) e^{\iu \nu_1 \tau_1} e^{-\iu \nu_2 \tau_2} e^{\iu \nu_3 \tau_3} e^{-\iu \nu_4 \tau_4},
    \label{eq:G2}
    \end{equation}
\end{widetext}
where the two-particle Green's function in terms of imaginary time ($\tau$) is defined as:
\begin{equation}
    G^{2}_{\substack{1234}}(\tau_1,\tau_2,\tau_3,\tau_4) \equiv \expec{\torder[\annihilation{1}(\tau_1) \creation{2} (\tau_2) \annihilation{3}(\tau_3) \creation{4}(\tau_4) ]}.
\end{equation}

Since our Hamiltonian Eq.~(\ref{eq:Hubbard_Model}) does not explicitly depend on time, one frequency can be removed by using energy conservation.
If not stated otherwise we will use the particle-hole (ph) notation:
\begin{align*}
    &\underline{\text{ph-notation:}} &    &  \underline{\text{pp-notation:}} \\
    & \nu_1 = \nu             &      & \nu_1 = \nu \\
    & \nu_2 = \nu - \omega            &      & \nu_2 = \omega - \nu' \\
    & \nu_3 = \nu' - \omega             &      & \nu_3 = \omega - \nu \\
    & \nu_4 = \nu'             &      & \nu_4 = \nu' \\
\end{align*}

The two-particle Green's function in Eq.~(\ref{eq:G2}) can be expressed in terms of disconnected (free) and connected (vertex) parts: 
    \begin{eqnarray}
         G^{(2),\omega \nu \nu' }_{\sigma \sigma'} &=& \delta_{\omega 0} G^{\nu}_{\sigma} G^{\nu'}_{\sigma'} - \delta_{\nu \nu'} \delta_{\sigma \sigma'} G^{\nu}_{\sigma} G^{\nu-\omega}_{\sigma'} 
\nonumber \\ &&       + \frac{1}{\beta}  G^{\nu}_{\sigma} G^{\nu-\omega}_{\sigma} F^{\omega \nu \nu'}_{\sigma \sigma'} G^{\nu' - \omega}_{\sigma'} G^{\nu'}_{\sigma'},
        \label{eq:G2_free_vertex}
    \end{eqnarray}
where $F$ is the so-called vertex function which encodes all scattering events on the
two-particle level and $G(\nu_1,\nu_2) = - \int_0^{\beta} \int_0^{\beta} d\tau_1
d\tau_2 \; \expec{\torder c_1(\tau_1) c^{\dagger}(\tau_2)} e^{\nu_1 \tau_1} e^{-\nu_2
\tau_2}$  is the
one-particle Green's function. 
All diagrams contained in the vertex function $F$ can be
classified unambiguously by the parquet decomposition
\begin{equation}
    F^{\omega\nu\nu'}_{\sigma\sigma'} = \Lambda^{\omega\nu\nu'} + \Phi^{\omega\nu\nu'}_{\text{pp},\sigma\sigma'} + \Phi^{\omega\nu\nu'}_{\overline{\text{ph}},\sigma\sigma'} + \Phi^{\omega\nu\nu'}_{\text{ph},\sigma\sigma'} \, .
\end{equation}
based on their two-particle reducibility, i.e., whether or not a diagram decomposes if one ``cuts'' two one-particle Green's function lines. For a diagrammatic depiction and the definition of $\Lambda$ and  $\Phi_r$ see  Fig.~\ref{fig:parquet_decomposition}.


\begin{figure*}[tb]
  \begin{minipage}{.64\textwidth}
    \includegraphics[width=\columnwidth]{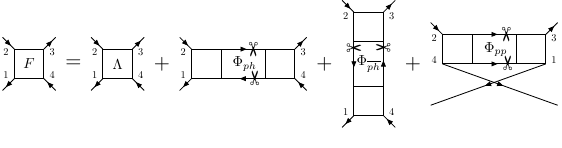}
  \end{minipage}\hfill
  \begin{minipage}{.34\textwidth}
\caption{Parquet decomposition of the full vertex function $F$ into its components based on the two-particle reducibility. $\Lambda$ is the fully irreducible vertex, and $\phi_{r}$ with $r=\{\ph,\phb,\pp\}$ is reducible part of $F$ in the particle-hole (ph), particle-hole transversal ($\phb$) and particle-particle ($\pp$) channel.}
\label{fig:parquet_decomposition}
\end{minipage}
\end{figure*}

 We now also define the irreducible vertices $\Gamma_{r} = F -\Phi_{r}$, where $r=\{\ph,\phb,\pp\}$ is either one of the three scattering channels. Furthermore, we assume SU(2) symmetry, i.e., restrict ourselves to the paramagnetic phase. This spin symmetry leads to only two independent spin combinations $F_{m} = F_{\uparrow \uparrow} - F_{\uparrow \downarrow}$ (magnetic) and $F_{d} = F_{\uparrow \uparrow} + F_{\uparrow \downarrow}$ (density) \cite{RMPVertex}. The vertex $F$ can also be expressed directly in terms of any irreducible vertex via the respective Bethe-Salpeter equation (BSE):
\begin{equation}
   F^{\omega\nu\nu'}_{d/m} = \Gamma^{\omega \nu \nu'}_{d/m,\text{ph}} + \frac{1}{\beta} \sum_{\nu_1} \Gamma^{\omega \nu \nu'}_{d/m,\text{ph}} G^{\nu_1} G^{\nu_1-\omega} F^{\omega\nu\nu'}_{d/m}.
    \label{eq:local_bse}
\end{equation}

Here, we have written the BSE only for $\Gamma_{\ph}$, and we use 
the inverse of Eq.~(\ref{eq:local_bse}) to extract the local $\Gamma_{\ph}$. To obtain a $q$-dependent susceptibility we use a BSE-like equation for the generalized, now momentum-$\mathbf k$ dependent susceptibility:

\begin{equation}
  \chi_{d/m}^{qkk'} =  \chi_{0}^{qkk'} - \chi_{0}^{qkk} \frac{1}{\beta^2} \sum_{k_1}  \Gamma^{qkk_1}_{\text{d/m}} \chi_{d/m}^{qk_1k'}.
\end{equation}
Here $k=({\mathbf k},\nu)$ and $q=({\mathbf q},\omega)$ are fermionic and bosonic four-point vectors in generalization of local, frequency-only-dependent quantities of the AIM.
Further, we approximate 
$\Gamma^{qkk_1}_{\text{d/m}} = \Gamma\freq_{\text{d/m}}$, i.e., the respective irreducible DMFT vertex. 
The physical susceptibility is finally obtained by summing over $k k'$, i.e. $\chi_{\text{phys}, d/m} = \frac{1}{\beta^2} \sum_{kk'} \chi_{d/m}^{qkk'}$. 
However, a susceptibility constructed this way contains divergences stemming from the mean-field phase transitions in DMFT, which largely overestimates critical temperatures~\cite{Rohringer2011}. This is even more true in two dimensions where the Mermin-Wagner theorem holds\cite{Mermin1966}. We remedy this artifact by employing a regularization parameter $\lambda$\cite{Katanin2009,RMPVertex} (instead of doing fully fledged parquet\cite{Valli2015} or self-consistent $\dga$ \cite{Kaufmann2020b}). Here, $\lambda$  is fixed by enforcing the sum-rule
\begin{equation}
 \frac{1}{2 \beta}\sum_{q} \big( \chi^{q, \lambda_m}_{m}  + \chi^{q, \lambda_d}_{d} \big)= \frac{n}{2} \cdot \big( 1 - \frac{n}{2} \big)
 \label{eq:chi_lambda} ,
\end{equation}
and $\chi^{q, \lambda}_{d/m} = \big[ \big( \chi^{q}_{d/m} \big)^{-1} + \lambda_{d/m} \big]^{-1}$ is the $\lambda$-corrected susceptibility.

%

\end{document}